\documentstyle[12pt]{article}
\begin{document}

\begin{center}
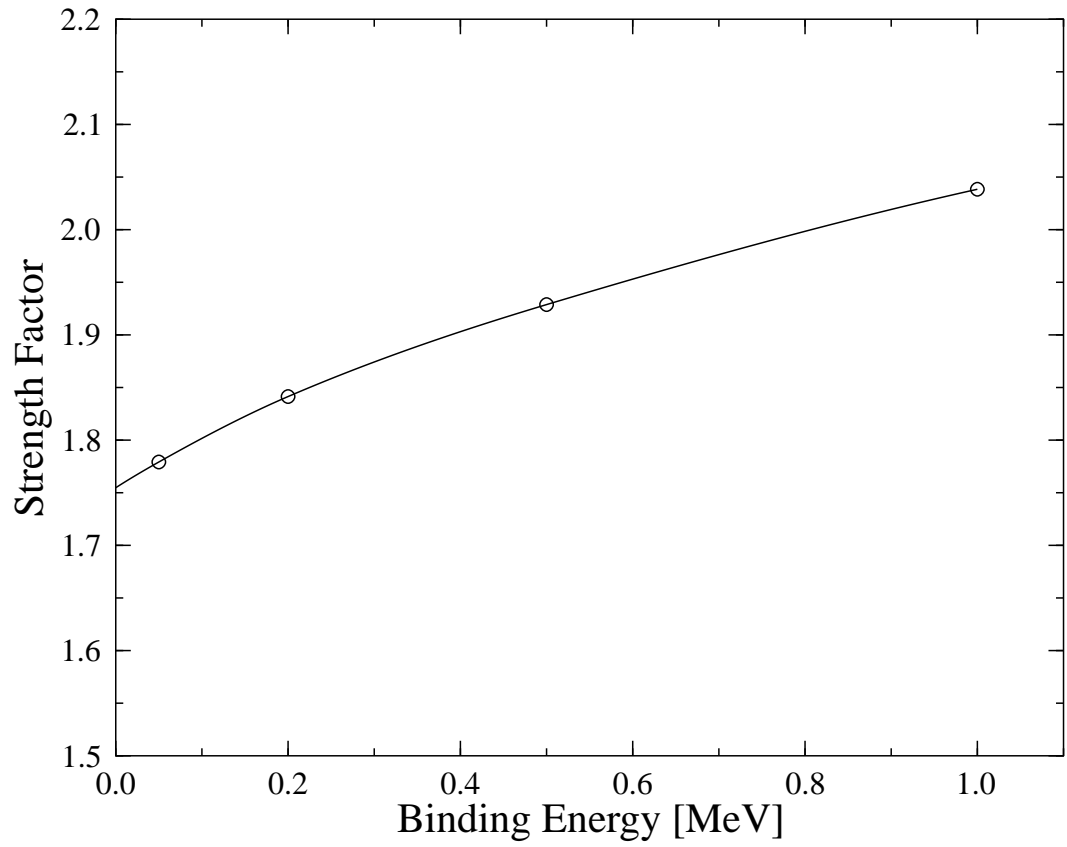
{\large Can the $\Sigma^- nn$ System be Bound?}\\
\end{center}
\vspace{.10in}
\begin{center}
A. Stadler\\
Department of Physics, College of William \& Mary\\
Williamsburg, Virginia 23185\\
\vspace{6pt}
and\\
\vspace{6pt}
B.F. Gibson\\
Theoretical Division, Los Alamos National Laboratory\\
Los Alamos, NM  87545\\
\end{center}

\vspace{.10in}
\begin{abstract}
Motivated by the $\Sigma$-hypernuclear states reported in
($K^-,\pi^{\pm}$) experiments, we have explored the possibility 
that there exists a particle-stable $\Sigma^- nn$ bound state.
For the J\"ulich \~A hyperon-nucleon, realistic-force model,
our calculations yield little reason to expect a positive-parity
bound state in either the $J = \frac{1}{2}$ or the $J = 
\frac{3}{2}$ channels.
\end{abstract}
%........1........2.........3.........4.........5.........6.........7........8
%\pagebreak
The question of the existence of $\Sigma$ hypernucei bound states 
--- narrow structure in hypernuclear spectra near the threshold 
for $\Sigma$ production in ($K^-,\pi$), {\it etc.}\ reactions ---
has intrigued physicists for more than a decade {\cite{wal88}}.  
The widths of such states were estimated to be rather broad 
($\sim$ 20 MeV) due to strong $\Sigma N \to \Lambda N$ 
conversion {\cite{gal80}, except in special cases.  Particularly 
interesting special cases are the maximum isospin few-body systems 
such as $\Sigma^- nn$, which cannot decay via $\Sigma N \to 
\Lambda N$ conversion because of charge conservation.  
However, the analysis by Dover and Gal {\cite{dov82}} of 
such maximum isospin states indicates that they are not expected 
to be the most bound.  They concluded, based upon the strong 
spin-isospin dependence of the $\Sigma N$ interaction, that 
the $T=0, J= \frac{1}{2}$ $\Sigma NN$ state should lie lowest
in energy --- lower than the two $T=1, J= \frac{1}{2}$ states or 
the $T=1, J= \frac{3}{2}$ and the $T=2, J= \frac{1}{2}$ 
configurations.  Unfortunately, the intrinsic width of the $T=0, 
J= \frac{1}{2}$ state was predicted to be much larger than the 
others.  Thus, it was not anticipated that narrow 
$\Sigma$-hypernuclear few-body states would be observed.

The interest in $\Sigma NN$ states was recently rekindled by the 
report of Hayano {\it et al} {\cite{hay89} that narrow structure 
was observed below the $\Sigma$ threshold in the stopping kaon 
reaction $^4$He($K^-,\pi^-$).  The structure in these data 
was confirmed by later inflight measurements {\cite{hay91}} and
is supported by earlier bubble chamber data {\cite{kat70}} for 
the exclusive $K^-$ $^4$He $\to \pi^- \Lambda p d$ reaction, which 
were recently reanalyzed {\cite{dal90}}.  This was surprising in 
view of the Dover and Gal analysis, in which the $T=\frac{1}{2}$ 
states were predicted to lie lower in energy but the 
$T=\frac{3}{2}$ states were predicted to have the narrower 
instrinsic widths.  Narrow structure was actually observed
in the $^4$He($K^-,\pi^-$) reaction below the threshold for 
$\Sigma$ production, whereas no evidence for an enhancement in 
that region was observed in the $^4$He($K^-,\pi^+$) spectra.  
The ($K^-,\pi^-$) reaction leads to both $T=\frac{1}{2}$ and $
T=\frac{3}{2}$ channels, while the ($K^-,\pi^+$) reaction leads 
only to the $T=\frac{3}{2}$ channel.  Thus, the observed 
structure was interpreted as a bound $^4_{\Sigma}$He 
hypernucleus with quantum numbers $T=\frac{1}{2}, J=0$.  
[The ($K^-,\pi$) spin-flip amplitude is small.]  In fact, 
Harada {\it et al.}\ {\cite{har87,har91}} had predicted such 
an $A=4$ bound state, based upon a central force approximation 
to the Nijmegen model D {\cite{nijD} hyperon-nucleon ($YN$)
potential.

Therefore, in spite of the theoretical analysis of Dover and 
Gal that suggests formation of a bound $\Sigma^- nn$ 
state is unlikely, one is led to ask whether state-of-the-art 
calculations based upon contemporary YN
potential models might indicate a possibility that the 
$T=2, J=\frac{1}{2}$ or $J=\frac{3}{2}$ states could be 
observed experimentally, either as a bound state in the 
continuum or as a three-body resonance.  \{Garcilazo 
{\cite{gar87}} argued on the basis of rank-one separable 
potentials that such a system is unbound.\}  One would prefer 
to explore all $\Sigma NN$ states, because $^3$He($K^-,
\pi^{\pm}$) experiments {\cite{bar92}} can excite only 
$T_z = \pm 1$ states.  [Target 
complications make the $^3$H($K^-,\pi^+$) reaction to the 
$T_z = -2$ state more difficult.]  However, including the 
$\Sigma N - \Lambda N$ coupling required by the $T=1$ states
leads to the technically difficult requirement that
one must solve the three-body equations for the continuum.
This has been accomplished for separable potentials
{\cite{afn93}, but not for local potential calculations.  For 
that reason we have confined our investigation to the possible 
existence of a $T=2$ bound state.

The Faddeev equations for the $\Sigma^- nn$ system were solved
in momentum space using the technical apparatus described in
% Ref.{\cite{haj83,sta92}}.  
Ref.{ \cite{sta91}}. 
The complication beyond standard triton calculations is that
the $\Sigma$ can be distinguished from the two neutrons, which
leads to a coupled pair of three-body equations instead 
of only the single equation that one finds for 
the comparable three-identical-particle problem. A more detailed
presentation of $YNN$ three-body bound-state equations can
be found in Ref.{ \cite{miy93}}.  
 
The baryon-baryon interactions are
assumed to act in all partial waves with $j \leq 1$ and with
positive parity.  This restriction yields a reasonable 
approximation to the converged binding energy in the 
three-nucleon system and can be expected to be  sufficient
for the purpose of determining whether a $\Sigma^- nn$ bound
state might exist. The effect of higher partial waves is 
certainly smaller than the variations induced by the use of
different baryon-baryon interaction models.  

Because we work in momentum space, we considered the 
J\"ulich {\cite{jueA} hyperon-nucleon interaction models.  In 
particular, we used the J\"ulich model \~A, an 
energy-independent one-boson-exchange approximation to the 
energy-dependent model A interaction.  The s-wave effective 
range parameters for $\Sigma^+ p$ scattering in these models 
are given in Table 1; we assumed equivalence for the $\Sigma^- 
n$ interaction for the purpose of this exercise.
We would point out, however, that the J\"ulich model differs 
qualitatively from the Nijmegen models {\cite{nijD,nijF,nijSC}}.  
The J\"ulich models are attractive for both spins, whereas 
the Nijmegen models exhibit a repulsive spin-triplet interaction.  
Thus, we have chosen the realistic $\Sigma N$ potential model 
that is most likely to support a $\Sigma^- nn$ bound state. For 
the $nn$ interaction we
employed the Nijmegen one-boson-exchange potential of 
Ref.{ \cite{nag78}}.

\begin{table}[bht]
\begin{flushleft}
{\bf Table 1.}  The $\Sigma^+ p$ scattering lengths and effective 
ranges in fm for the J\"ulich potential models listed
\end{flushleft}
\begin{tabular}{clcc|ccccccccc} \hline
\hspace*{1cm}& Model & Ref. &\hspace{1cm}&\hspace{2cm}& a$^s$ &
  \hspace{.5cm}& r$^s_o$ &\hspace{2cm}& a$^t$ &\hspace{.5cm}& r$^t_o $ 
  &\hspace{1cm} \\ \hline
 & & & &  & & & & & & & & \\
\hspace*{1cm}& J\"ulich A & {\cite{jueA}}&\hspace{1cm}&\hspace{2cm}&
 -2.28 &\hspace{.5cm}& 4.96 &\hspace{2cm}& -0.76 &\hspace{.5cm}& 2.50  
  &\hspace{1cm} \\
 & & & &  & & & & & & & & \\
\hspace*{1cm}& J\"ulich \~A&{\cite{jueA}}&\hspace{1cm}&\hspace{2cm}&
 -2.26 &\hspace{.5cm}& 5.22 &\hspace{2cm}& -0.76 &\hspace{.5cm}& 0.78  
  &\hspace{1cm} \\
 & & & &  & & & & & & & & \\
 \hline
\end{tabular}
\end{table}

Our search for a bound $\Sigma^- nn$ system with $J^{\pi} = 
\frac{1}{2} ^+$
proved negative.  In retrospect this is not surprising in view of 
the fact that the spin-singlet $\Sigma N$ interaction is the stronger, 
whereas the spin-triplet potential dominates: the average interaction 
is $\frac{1}{4}$V$^s$ + $\frac{3}{4}$V$^t$.  \{Lack of binding was
also found for the hypertriton using the $T=\frac{1}{2}$
$\Lambda N - \Sigma N$ potentials of this same J\"ulich \~A model
{\cite{miy93}}.~\}
To understand how far away a resonance might lie, we have multiplied 
the total interaction by a variable factor, increasing that factor 
until binding was achieved.  A plot of the strength factor versus 
the binding energy obtained is shown in Fig.\ 1.  Because the 
factor needed to produce binding is greater than 1.7, we do not 
expect any low-lying resonance in the $\Sigma^- nn$ system.  In the 
$J^{\pi} = \frac{3}{2} ^+$ case, a spectator $\Sigma^-$ must be at 
least in a p-wave relative to the $nn$ pair in order to reach spin-3/2, 
because s-wave neutrons will necessarily be paired to spin-0.  
Therefore, it was expected that the $J = \frac{3}{2}$ state will be 
unbound in view of the finding that there is no $J = \frac{1}{2}$ 
bound state.  Indeed, that was the case.

\vspace{12pt}
\noindent {\bf Acknowledgement}
\vspace{4pt}

The work of BFG was performed under the auspices of the U.\ S.\
Department of Energy. The work of AS was supported by the DOE 
under Grant No.\ DE-FG05-88ER40435. The calculations were performed
at the National Energy Research Supercomputer Center (Livermore).
AS wishes to thank the CEBAF Theory Group and the T-5 Theory group
for their hospitality when part of this work was performed.

\pagebreak

\begin{figure}
\vspace*{10cm}
\begin{flushleft}
{\bf Fig.\ 1} Strength factor by which the total $\Sigma N$
interaction is multiplied versus the $\Sigma^- nn$ binding
energy. The circles represent the actual
calculations, the solid line is drawn to guide the eye.
\end{flushleft}
\end{figure}

\end{document}